\documentclass[10pt]{article}

\usepackage{epsfig,bm,titlesec}
\usepackage{hyperref}
\usepackage{graphicx}

\newcommand{\NPA}[3]{#3 Nucl.\ Phys.\ {\bf A#1} #2}

\newcommand{\PLB}[3]{#3 Phys.\ Lett.\ B\ {\bf #1}  #2}

\newcommand{\PRC}[3]{#3 Phys.\ Rev.\ C\ {\bf #1} #2}
\newcommand{\PRD}[3]{#3 Phys.\ Rev.\ D\ {\bf #1}  #2}

\begin{document}

\title{Constraint on the magnetic field for the stable strange quark matter}

\author{
 Xin-Jian Wen,
Xiao-Wen Jiang \\
\small Institute of Theoretical Physics, Shanxi University, Taiyuan
030006, China
      }

\maketitle

\begin{abstract}

The quasiparticle model is employed to investigate the quark matter
at finite chemical potential. The effective bag constant is derived
to be dependent on both the chemical potential and the magnetic
field. The self-consistent thermodynamics is fulfilled that the free
energy minimum corresponds to the zero pressure. It is shown that
the strong magnetic field is helpful for the stabilization of the
strange quark matter. However, the increase in the coupling constant
and the vacuum bag constant could reduce the stability. For the
absolutely stable strange quark matter, there is a lower limit of
the allowed magnetic field, which rises with the increase in the
coupling constant and the vacuum bag constant.
\end{abstract}


\maketitle

\section{Introduction}
\label{sec:intro}

It is well known that the interaction of quarks and gluons is
described by quantum chromodynamics (QCD). Due to the conversion of
u-, d-quarks into s-quarks by the weak reactions, strange quark
matter (SQM) could be more stable than hadronic matter. SQM is
suggested to occur in extreme hot and high density condition at
relativistic heavy ion collision experiment or in neutron stars.
With growing evidence, strongly interacting Quark Gluon Plasma is
realized near $T_c$ in relativistic heavy-ion collisions
\cite{Shur04,Gyul05}. A first order transition is expected to happen
from the cold nuclear matter to quark matter. However, the location
of the phase transition to deconfined quark matter remain poorly
known to this day. In fact, the transition can be accounted for by
the stability of the quark matter with respect to nuclear matter at
zero temperature and the structure of condense stars
\cite{Kettner:1994zs,Dexheimer:2013eua}. Furthermore, this is
related to the possibility of the entire stars being made up of
self-bound quark matter
\cite{Peng:2018qlh,Felipe:2007vb,Liu:2023kzy}.

The cold dense quark matter is investigated in MIT bag model dating
back 50 years \cite{Chodos:1974pn}. The interaction between quarks
is represented by a phenomenological bag constant and produces a bag
pressure to the non-interacting system \cite{Alcock,Chodos:1974je}.
40 years ago, Bodmer-Witten hypothesis suggested that clusters of
SQM can be more stable than the most stable atomic nucleus. The
possibility that dark matter could be composed of macroscopic
clusters of SQM would rely on the stability of SQM with a suitable
confinement. The first order transition seemed to not be a crucial
requirement in order to produce large clusters of quarks
\cite{DiClemente:2024lzi}. In the bag model, the quark mass is
infinitely large outside while it is constant within the bag. As is
well known in nuclear physics, particle masses vary with
environment. Taking advantage of the medium effect, the strong
interactions between the elementary degrees of freedom are
incorporated through the medium-dependent quasiparticle mass
\cite{Peshier:1994zf,Ma:2023stj}. The quark confinement can be
described by the density dependence of quark masses
\cite{Fowler:1981rp,Wen:2010zz}. The quasiparticle model with a
chemical potential and/or temperature dependent quark mass is
considered to be superior to the plain bag model. Peshier proposed
that the thermal mass depends on the perburbative expression for
plasma frequencies \cite{Peshier1996,Peshier:1999ww}. Bannur et.al.
improved the model by the density dependent expression and the full
HTL dispersion relation instead of the approximate dispersion
relation \cite{Bannur08}. In Leonidov's approach, the entropy and
baryon number conservation is maintained with effective bag pressure
in the deconfinement phase transition
\cite{Leonidov:1993it,Joshi:2020lwn}. Under the density and/or
temperature dependent mass scale, the vacuum energy is described by
effective bag function, which is crucial to the thermodynamical
consistence \cite{Thaler:2003uz}. At zero temperature, the quark
mass depends merely on chemical potentials
\cite{Schertler:1996tq,Wen:2009zza}. It will be more convenient to
investigate the self-consistent thermodynamics and pave the way for
further work with such models. The aim of this paper is to
demonstrate the magnetic field effect on the equation of state at
finite chemical potentials.

This paper is organized as follows. In Section \ref{sec:model}, we
present the self-consistent thermodynamics of the quasiparticle
model in a strong magnetic field. The effective bag constant is
derived to reflect confinement scheme. In Section \ref{sec:result},
the numerical results are shown at the finite density in a strong
magnetic field. The detailed discussions are focused on the
stability of SQM. The last section is a short summary.

\section{Thermodynamics in quasiparticle model}\label{sec:model}
The quark-gluon plasma phase is a new state of matter consisting of
three flavor quarks. The whole understanding of the QCD phase should
be based on the properties of quark matter surrounded by the medium
at finite temperatures and densities. The important feature of the
medium dependence is quark effective masses influenced by QCD
nonperturbative properties. In a simple case, the medium is
characterized by a whole se of collective quasiparticles in the
vacuum. For the medium dependence of the quark quasiparticle model,
the effective quark mass $m_i^*$ is derived at the zero momentum
limit of the dispersion relation following from the effective quark
propagator by resuming one-loop self energy diagrams in the hard
dense loop (HDL) approximation \cite{Schertler:1996tq}. The
in-medium effective mass of quarks can thus be expressed as
\cite{Schertler:1996tq,Schertler:1997vv,Peshier:1999ww,Bannur:2007tk,Wen:2009zza}
\begin{equation}\label{mass}
m_i^*(\mu_i)=\frac{m_i}{2}+\sqrt{\frac{m_i^2}{4}+\frac{g^2\mu_i^2}{6\pi^2}},
\ \ (i=u,d,s),
\end{equation} where $m_i$ is the current mass of corresponding
quarks and the constant $g$ is related to the strong interaction
constant $\alpha_s$ by the equation $g=\sqrt{4\pi \alpha_s}$. The
quasi-particle idea can be recalled backward to the work by Fowler
et.al.\cite{Fowler:1981rp} that the particle mass may change with
the environment parameters. The effective quark mass $m_i^*$
increases with $g$ at the fixed quark chemical potential $\mu_i$.
The unknown coupling $g$ is to be determined as a decreasing
function of the temperature and density reflecting the asymptotic
freedom of QCD.

The quasiparticle contribution to the thermodynamic potential
density is given as .
\begin{eqnarray}\label{thero-int}
\Omega_i &=& \frac{d_i |q_iB_m|}{2\pi^2}
\sum_{\nu=0}^{\nu_i^\mathrm{max}} (2-\delta_{\nu 0})
  \int_{0}^{p_F}
    \left(
E_i-\mu_i
    \right) \mbox{d}p_z\, ,
\end{eqnarray}
where the degeneracy factor $d_i=3$ for the color freedom of
$i$-flavor quarks. The single particle energy eigenvalue
$E_i=\sqrt{{m_i^*}^2+p_z^2+ 2 \nu_i |q_i B_m|}$ sensitively depends
on the magnetic fields, the thermal mass $m^*_i$ and $z$-component
momentum $p_z$. At zero temperature, the upper limit
$\nu_i^\mathrm{max}$ of the summation index $\nu_i$ can be
understood from the positive value requirement on Fermi momentum and
is defined by $\nu_i^\mathrm{max}=(\mu_i^2-{m_i^*}^2)/(2 |q_i
B_m|)$. In order to include the vacuum contribution, the total
pressure should be a sum of a matter pressure and a bag constant
$P(\mu_i)=-\Omega_i-B(\mu_i)$. The $B(\mu_i)$ represents the model
of confinement used in our studies of SQM.

The pressure function would include the contribution of the matter
pressure and density-dependent bag pressure, namely, $P=-\Omega-B$.
If the thermodynamic potential depends on the state variable
implicitly via phenomenological parameters $m_i^*$, the
corresponding stationarity condition should be required as
\cite{Gorenstein:1995vm}
\begin{eqnarray} \left.\frac{\partial P}{\partial
m_i^*}\right |_{T,\mu_i}=0. \label{eq:condition}
\end{eqnarray}
which has been widely employed in the quasiparticel model at finite
temperature and density
\cite{Peshier:1999ww,Schneider:2001nf,Bannur:2006js,Giacosa:2010vz}.
The condition respects the chiral symmetry restoration in the plasma
\cite{Peshier:1998ei}. At zero temperature, we get the bag function
through the integral
\begin{eqnarray}\label{Bexp}
B_i^*(\mu_i)&=&- \frac{d_i |q_iB_m|}{2\pi^2}
\sum_{\nu=0}^{\nu_i^\mathrm{max}} (2-\delta_{\nu 0}) \int^{\mu_i}_0
\int_{0}^{p_F} \frac{m_i^*}{E_i}
 \frac{\mathrm{d}
m^*_i}{\mathrm{d}\mu_i}\mathrm{d}p_z \mathrm{d}\mu_i,
\end{eqnarray}
where the allowed chemical potential in the outer integral should
start from zero. Specially, the equality $\mu_i=m_i^*$ would lead to
the vanishing Fermi momentum and the zero bag function. The upper
limit of the inner integral is the Fermi momentum dependent on
$\mu_i$. So the total bag pressure is a sum of the flavor and vacuum
contributions as $B=\sum_i B^*_i(\mu_i)+B_0$. For a stable SQM, the
proposed value of the Bag constant should be within the following
range $(55 \sim 75)$ MeV/fm$^3$ \cite{Alcock}. Compared with the
standard Statistical Mechanics (SM), one can recover the
thermodynamics consistency of system $T$-$\mu$-$B_m$- dependent
Hamiltonian with the extra term $B(T,\mu,B_m)$. The meaning of $B$
plays an important role in the physical conclusion. The
interpretation of $B$ was first given by Gorenstein and Yang in
Ref.\cite{Goren1995}. In quasiparticle model, $B$ is the system
energy in the absence of quasi-particle excitations, which cannot be
discarded from the energy spectrum \cite{Gardim07}. In this sense,
$B$ plays as bag energy or bag pressure through application in
bag-like model. One can interpret the confinement mechanism
considering $B$ as the difference of perturbative vacuum and
physical vacuum \cite{Patra1996}. The bag constant is taken as a
critical parameter in the determination of whether the ground state
of baryon matter in QCD is the ordinary nucleus or a quark matter
state \cite{arXiv:2410.19678} In literature, many work have been
devoted to the description of the bag constant.

The number density will not feel the influence of the bag constant,
because of the followings \cite{Wen:2009zza},
\begin{eqnarray} n_i=\frac{\partial P}{\partial
\mu}+ \frac{\partial P}{\partial m^*_i}\frac{\partial
m^*_i}{\partial \mu_i} = \frac{\partial P}{\partial \mu},
\end{eqnarray} where the second partial derivative vanishes due to the
condition (\ref{eq:condition}). The essential thermodynamic relation
will not change with the inclusion of the quark mass scale and the
density-dependent bag constant. We can give the number density by
the sum of landau levels,
\begin{eqnarray} n_i=\frac{d_i |q_iB_m|}{2\pi^2}
\sum_{\nu=0}^{\nu_i^\mathrm{max}} (2-\delta_{\nu 0})
\sqrt{\mu_i^2-{m_i^*}^2- 2\nu |q_iB_m|}.
\end{eqnarray}

It is well known to us that the asymptotic freedom is an essential
feature of QCD. In literature, it can be included in effective
chiral quark models by the running coupling constant. The
approximate expression for the running quantity $g(\mu)$ was
proposed as a decreasing function of chemical potential in
Ref.~\cite{Patra1996}.  Based on the fact of the magnetic catalysis
of chiral symmetry breaking that incorporates the effects of the
field in the running coupling, we take the following expression,
\begin{eqnarray}g^2(\mu, B_m)=\frac{48 \pi^2}{(33-2
N_f)\ln(0.8\frac{\mu^2}{\Lambda_1^2})[1+\alpha\ln(1+\beta\frac{eB_m}{\Lambda_2^2}
)]}
\end{eqnarray}
where the scale-fixing parameters are $\Lambda_1=120$ MeV, and
$\Lambda_2=200$ MeV in our work. The free parameters are adopted as
$\alpha=2$ and $\beta=0.000327$ \cite{Farias:2014eca}. Accordingly,
the derivative of the mass with respect to the chemical potential
$dm^*_i/d\mu_i$ in Eq. (\ref{Bexp}) is modified as,
\begin{eqnarray}\label{eq:dmdu}
\frac{dm^*_i}{d\mu_i}=\frac{\partial
m^*_i}{\partial\mu_i}+\frac{\partial m^*_i}{\partial
g}\frac{dg}{d\mu_i}.
\end{eqnarray}The running coupling constant changes the slope of the
mass $m^*$ with respect to the quark chemical potential.
\section{Numerical result}\label{sec:result}
\begin{figure}
\centering
\includegraphics[width=8.cm,height=6cm]{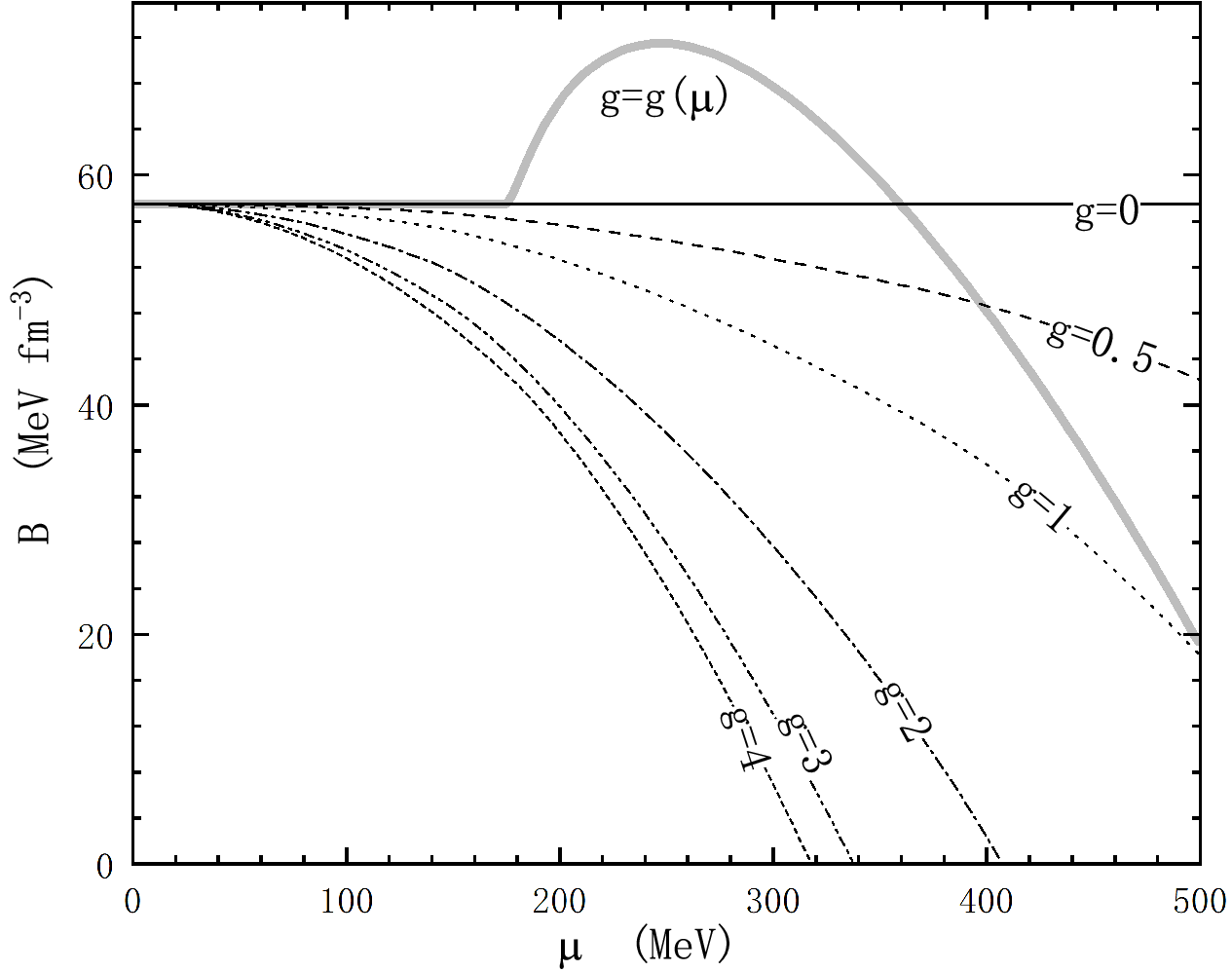}
\caption{ The variation of the bag constant as a decreasing function
of the quark chemical potential $\mu$ is shown at the magnetic field
$eB_m=0.2$ GeV$^2$ and different coupling constants $g=0$, 0.5, 1,
2, 3, and 4. On the top $g=0$ produces the conventional bag
constant. }\label{fig1}
\end{figure}
It is known to us that the low limit of the bag constant is required
for the assumption of the stable quark matter. A lower value
$B^{1/4}=76.8$ MeV is estimated from GMOR \cite{arXiv:2410.19678}.
On the other hand, the equation of state with a larger value
$B^{1/4}=174.2$ MeV is possible to mimic the mass-radius behavior of
nucleonic stars \cite{Alford:2004pf}. To form stars with a larger
mass, it was proposed to set the bag constant to a low value. The
most probable range of values $B^{1/4}\cong 150\sim 200 $MeV is
proposed to be consistent with the estimate from hadron spectroscopy
\cite{Kouno:1988bi}. In Figure 1, the density dependent bag constant
is shown as a decreasing function the chemical potential at the
fixed coupling constant. At larger coupling constant, the bag
constant would more rapidly drop down with a larger slope. At $g=0$,
the medium effect is vanished and we get the conventional bag model
with a fixed bag constant $B_0^{1/4}=145$ MeV. As the coupling
constant increases in the figure, the bag constant seems to have
lower value. Under the running coupling $g(\mu)$, the effective bag
constant is evaluated  by submitting Eq.(\ref{eq:dmdu}) into Eq.
(\ref{Bexp}). At lower chemical potential, the vacuum bag constant
is hold due to the larger mass resulted by the larger value
$g(\mu)$. A sudden increase occurs only for the chemical potential
larger than the quark mass. The non-monotonous behavior is produced
by the parameterization of the running coupling. In the region of
high chemical potential, the bag constant for different couplings
will have common trend. A natural question to ask is: how strong
coupling constant is permitted for the stable state? We will give
the answer by the following analysis.

\begin{figure}
\centering
\includegraphics[width=8.cm,height=6cm]{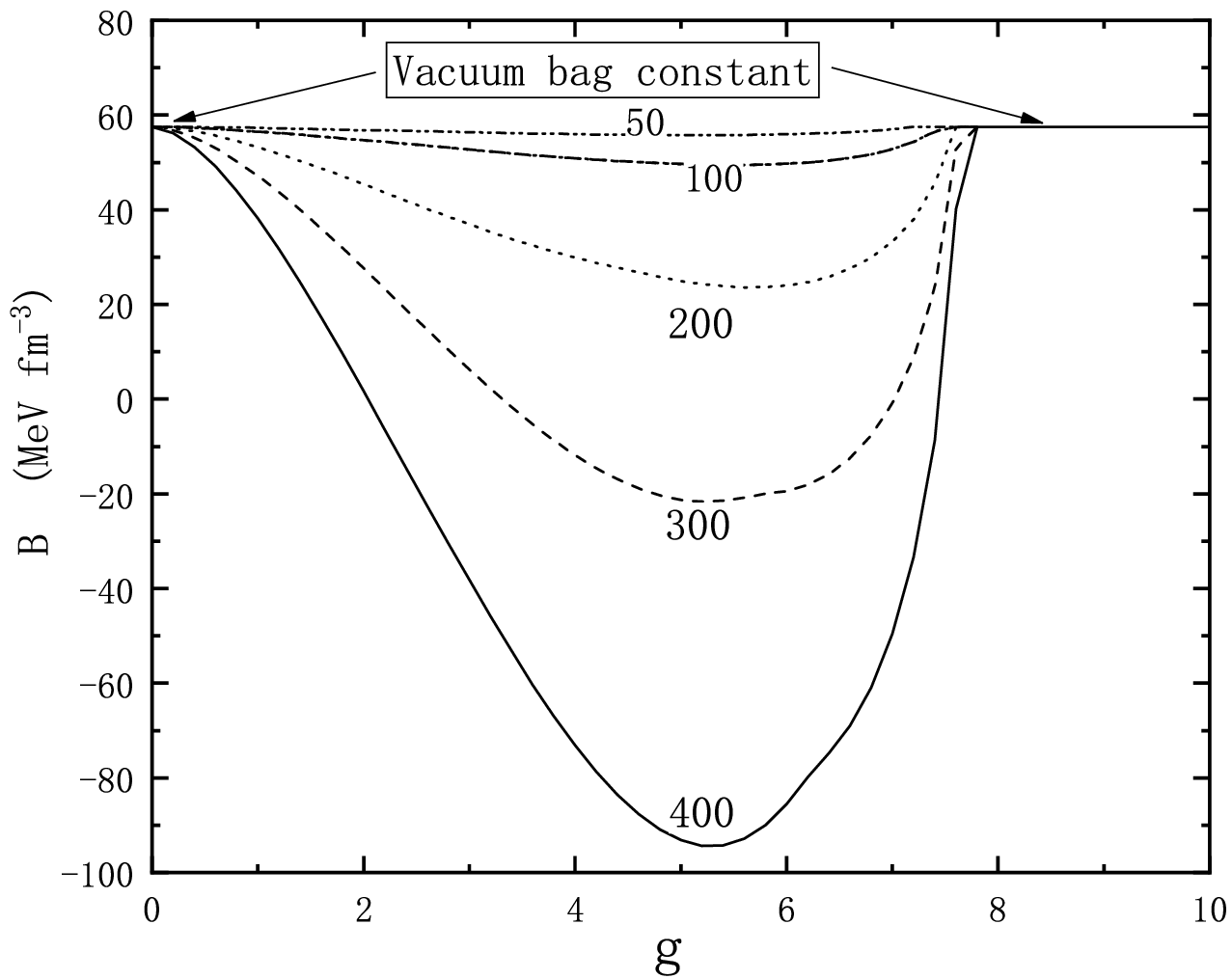}
\caption{ The variation of the bag constant with the coupling
constant $g$ is shown at $eB_m=0.2$ GeV$^2$. The line on the top at
$\mu=50$ MeV is very close to the vacuum bag constnat $B_0=58
$MeV$\cdot$fm$^{-3}$. }\label{fig2}
\end{figure}
In Fig. \ref{fig2}, the bag constant is shown as a non-monotonic
function of the coupling constant at fixed chemical potential. The
bag constant would return to the vacuum constant $B_0$ at larger
coupling for a given chemical potential. At the coupling constant
larger than $\sqrt{6}\pi$, the particle with larger mass would have
zero number density $n=0$ in vacuum state, which is consistent with
the density-density parameterization of bag constant
$B(n)=B_\infty+(B_0-B_\infty)\exp(-\gamma(n/n_0)^2)$ in the work of
Burgio \cite{Burgio02}. In the context of compact stars we are not
interested, the chemical potential is large and the temperature
practically zero. At larger chemical potential $\mu=500$ MeV, the
medium effect becomes so evident that the bag constant is suppressed
at an intermediate coupling constant.

\begin{figure}
\centering
\includegraphics[width=8.cm,height=6cm]{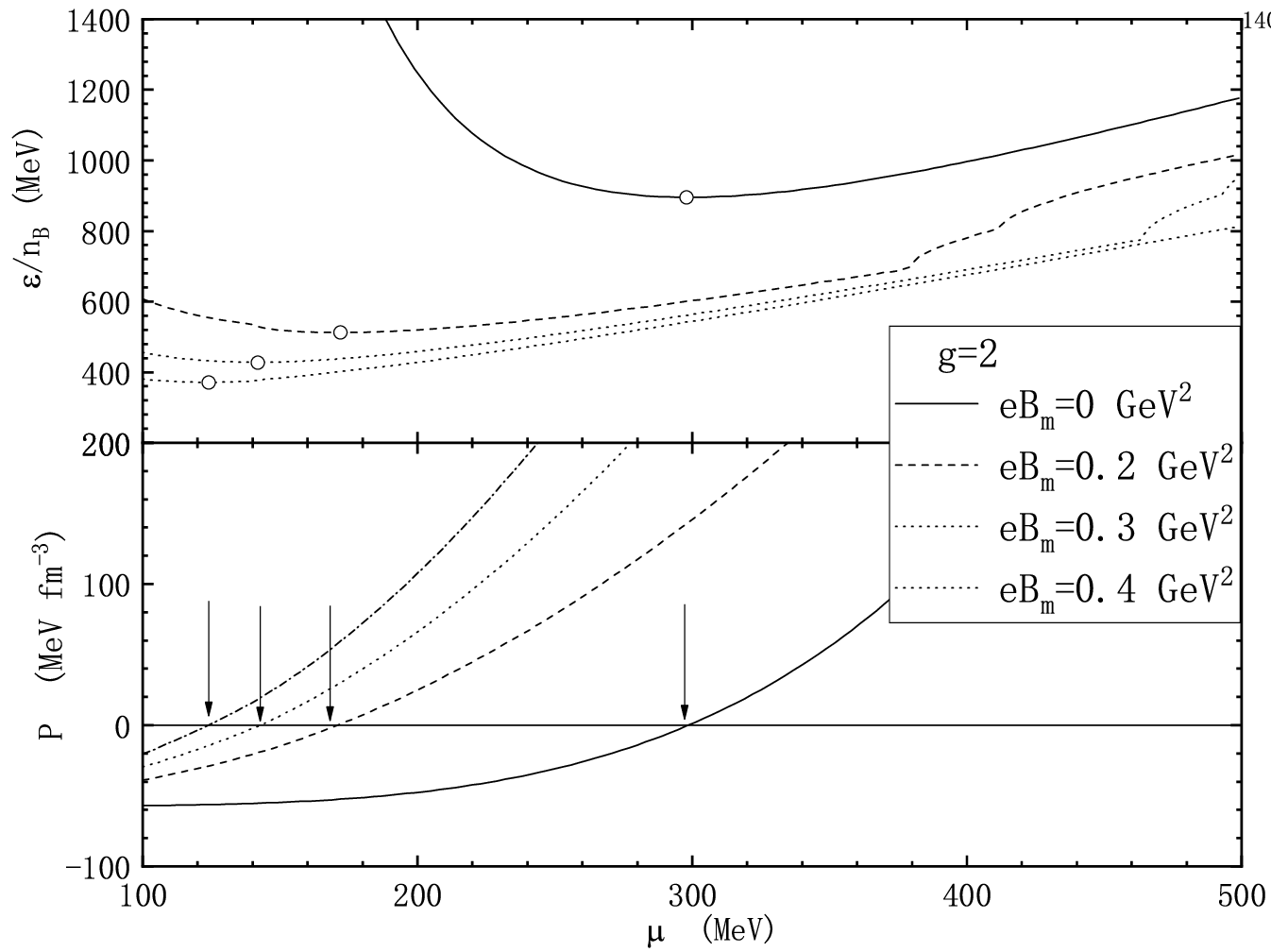}
\caption{ The free energy per baryon on the top panel and the
pressure function on the bottom panel are shown as a function of
chemical potential at the magnetic field $eB_m=0$, 0.2, 0.3, and 0.4
GeV$^2$. }\label{fig3}
\end{figure}

The free energy density is defined as $\epsilon=\Omega+B+\mu \sum_i
n_i $, which will include the bag constant as in the pressure. The
free energy per baryon has a minimum value $3 \mu$ at the zero
pressure. In Fig. \ref{fig3}, the free energy per baryon and the
pressure function are shown with and without the strong magnetic
field at the coupling constant $g=2$. For stable quark matter, the
matter pressure must balance the bag pressure $-B$ at the quark
chemical potential smaller than the critical value
$\mu_c=\mu_B/3=310$ MeV. It is shown that the minimum free energy
per baryon marked by the empty circle occurs exactly at the zero
pressure indicated by the arrows. Furthermore, the zero pressure
position moves far smaller than the $\mu_c$ with the increase in the
magnetic field. It can be concluded that the stability of the SQM is
absolutely enhanced by the strong magnetic field.

\begin{figure}
\centering
\includegraphics[width=8.cm,height=6cm]{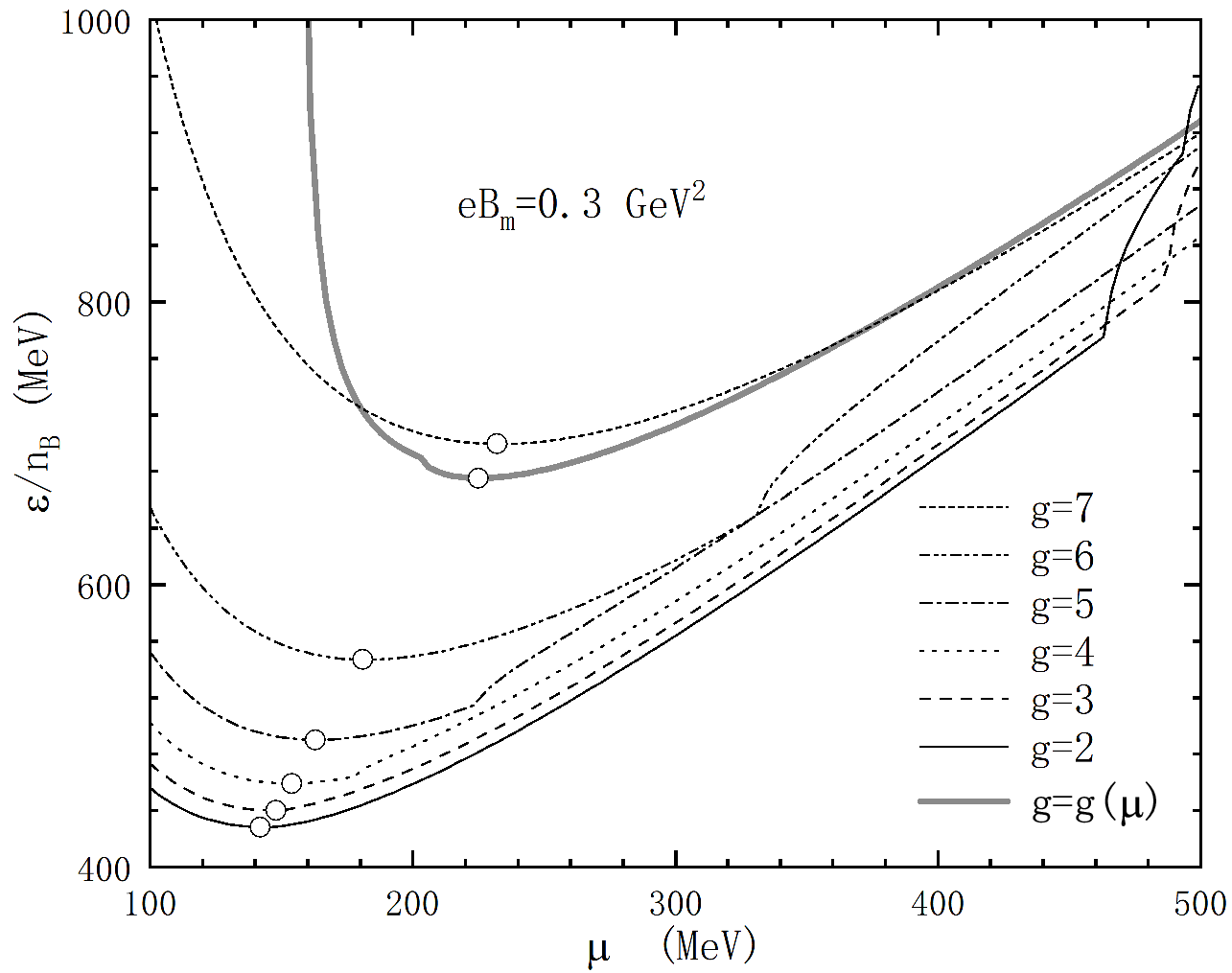}
\caption{ The free energy per baryon as a function of chemical
potential is enhanced by an increase in the coupling constant at the
magnetic field $eB_m=0.3$ GeV$^2$. }\label{fig4}
\end{figure}

In Fig. \ref{fig4}, the free energy per baryon is shown at the fixed
magnetic field $eB_m=0.3$ GeV$^2$. The minimum free energy per
baryon is marked by the empty circle, which is zero pressure point.
As the coupling constant increases, the $\epsilon/n_B$ is becoming
large, and therefore the stability would be reduced. By comparison
with the magnetic field in Fig.~\ref{fig3}, the coupling interaction
has an opposite effect on the stability of SQM. It should be
mentioned that the vacuum bag is adopted as $B_0^{1/4}=145$ MeV. As
long as the chemical potential position $``\circ"$ is located on the
left of $\mu_c=310$ MeV, the SQM is absolutely stable with respect
to ordinary nuclear matter. The $\epsilon/n_B$ for the running
coupling $g(\mu)$ marked by the grey line is higher than other cases
of $g<6$ due to its larger value of the bag constant. Moreover, the
slope of $\epsilon/n_B$  on the left branch of $\mu<\mu_\circ$ is
more steep for the running coupling $g(\mu)$ becomes much larger
than the fixed coupling constant.

\begin{figure}
\centering
\includegraphics[width=8.cm,height=6cm]{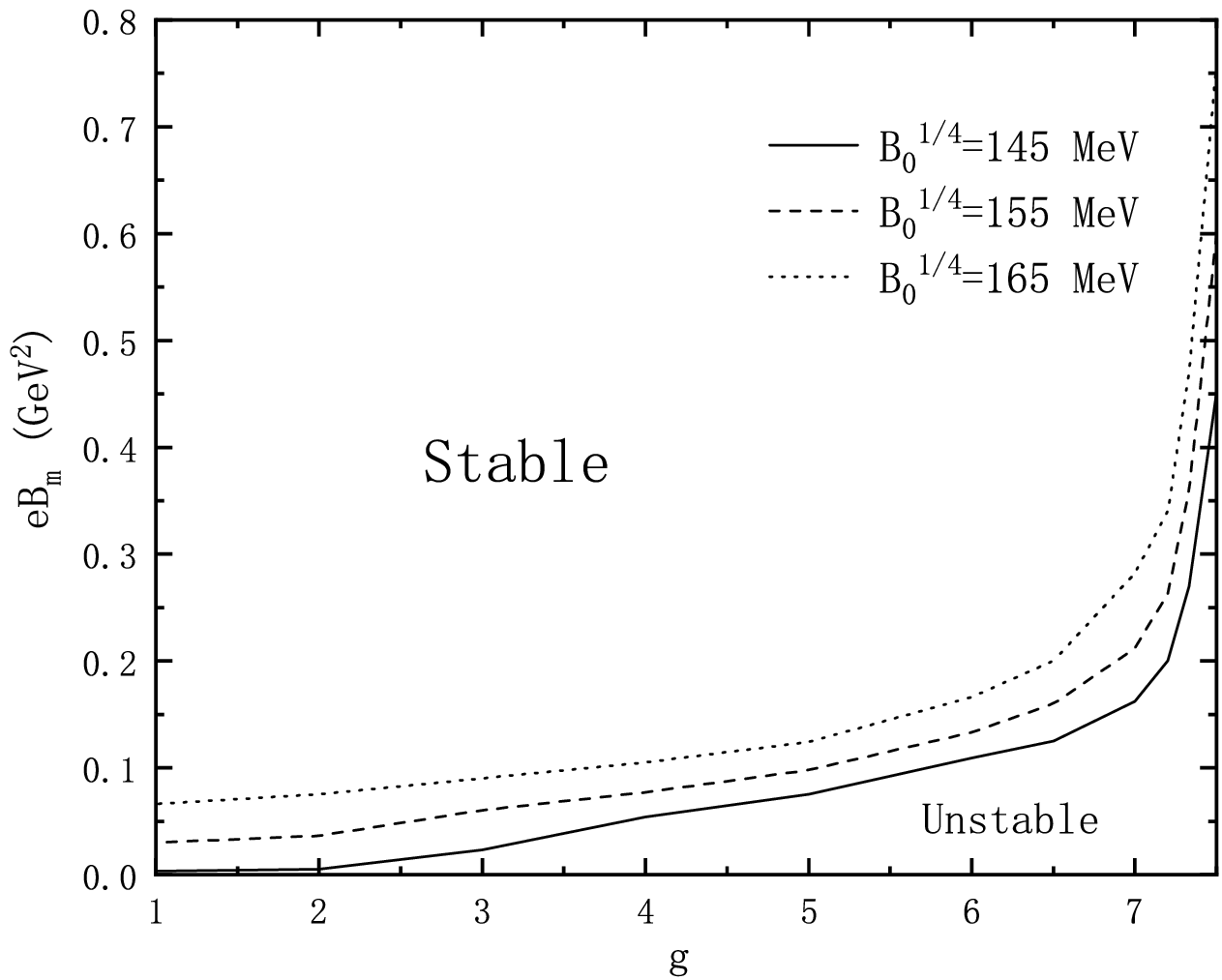}
\caption{ The stability window in the $eB_m$-$g$ plane is shown at
different vacuum bag constants $B_0$. }\label{fig5}
\end{figure}

Since the coupling constant and the magnetic field have opposite
effect on the stability of SQM, the minimum magnetic field would be
changed by the increase in the coupling constant. The stability of
the SQM can be described by the free energy per baryon smaller than
the iron nuclei $930$ MeV. In Fig. \ref{fig5}, the stability window
is shown in the $eB_m$-$g$ plane for different vacuum bag constants
$B_0$. The line $\epsilon/n_B=930$ MeV marks the estimated lower
limit of the range of the strong magnetic field. The stable region
would shrink as the vacuum bag constant $B_0$ increases, and the
boundary of the region would move to the stronger magnetic field and
smaller coupling constant. In other words, the weak coupling
constant is favored at a fixed magnetic field. For much larger
vacuum bag constant $B_0$, the stronger magnetic field is required
for the existence of the stable SQM.

\section{Conclusion} \label{Sec:conls}

In this paper the quasiparticle model has been employed to
investigate the stability of SQM in a strong magnetic field. We have
derived the effective bag constant self-consistently instead of the
ansatz to describe the density dependent bag for the confinement
scheme. The effective bag constant has been shown as a decreasing
function of the chemical potential. However, as the increase in the
coupling constant, the bag constant displays a non-monotonous
behavior, which approaches the vacuum bag constant $B_0$ at both the
zero coupling and the much larger coupling constant. The stability
of SQM has been reflected by the free energy per baryon. The
thermodynamically self-consistence is fulfilled that minimum free
energy per baryon occurs at the zero pressure. Finally, we have
found that the increase in magnetic field would enhance the
stability of quark matter, even though the larger coupling constant
and the vacuum bag would reduce its stability. The lower limit of
the magnetic field would rise as the coupling constant becomes
larger. A realistic calculation of the magnetic field profile inside
strongly magnetized neutron stars is reported in Ref.
\cite{Dexheimer:2016yqu}. They suggested that the magnetic fields
increase relativistic slowly with increasing baryon chemical
potential  of magnetized matter. So there is a strong possibility
that the SQM exists in the core of neutron stars. We hope our work
could provide a simple method to the constraint on the magnetic
field strength in the SQM, which could be the consistent of the
massive compact object \cite{Bianchi:2024xnf}.

\Large\begin{flushleft} {\bf
Acknowledgements}\end{flushleft}\normalsize

The authors would like to thank the National Natural Science
Foundation of China for support through Grants No.11875181 and
No.12047571. This work was also sponsored by the Fund for Shanxi
“1331 Project” Key Subjects Construction.

\end{document}